\documentclass{article}

\usepackage{arxiv}

\usepackage[utf8]{inputenc} 
\usepackage[T1]{fontenc}    
\usepackage{hyperref}       
\hypersetup{    
    colorlinks,
    citecolor = black,
    filecolor = black,
    linkcolor = black,
    urlcolor = black
}
\usepackage{url}            
\usepackage{booktabs}       
\usepackage{amsfonts}       
\usepackage{nicefrac}       
\usepackage{microtype}      
\usepackage{lipsum}		    
\usepackage{graphicx}
\usepackage[round]{natbib}
\usepackage{doi}
\usepackage{todonotes}
\usepackage{multicol, caption}
\newenvironment{Figure}
  {\par\medskip\noindent\minipage{\linewidth}}
  {\endminipage\par\medskip}
  
\title{Reinforcement Learning and Graph Neural Networks for Probabilistic Risk Assessment}

\author{ \href{https://orcid.org/0009-0008-5069-1547}{\includegraphics[scale=0.06]{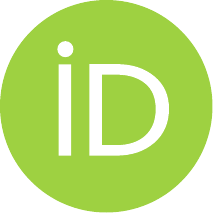}\hspace{1mm}Joachim~Grimstad}, \href{https://orcid.org/0000-0002-6935-3563}{\includegraphics[scale=0.06]{orcid.pdf}\hspace{1mm}Andrey~Morozov} \\
	Institute of Industrial Automation and Software \\ Engineering,
	University of Stuttgart\\
	Pfaffenwaldring 47, 70550 Stuttgart, Germany\\
	\texttt{\{first.last\}@ias.uni-stuttgart.de} \\
}

\renewcommand{\shorttitle}{Reinforcement Learning and Graph Neural Networks for Probabilistic Risk Assessment}

\begin{document}
\maketitle

\begin{abstract}
This paper presents a new approach to the solution of Probabilistic Risk Assessment (PRA) models using the combination of Reinforcement Learning (RL) and Graph Neural Networks (GNN). The paper introduces and demonstrates the concept using one of the most popular PRA models - Fault Trees. This paper's original idea is to apply RL algorithms to solve a PRA model represented with a graph model. Given enough training data, or through RL, such an approach helps train generic PRA solvers that can optimize and partially substitute classical PRA solvers that are based on existing formal methods. Such an approach helps to solve the problem of the fast-growing complexity of PRA models of modern technical systems.

\end{abstract}




\begin{multicols}{2}

\section{Introduction}\label{sec: Introduction}

In the era of relentless technological advancement, systems across various domains are evolving at an unprecedented pace, becoming increasingly complex. This increase in complexity poses significant challenges for engineers, regulators, and policymakers. Despite the advancements in technology, these complex systems remain susceptible to faults and failures, which can pose serious safety risks if not adequately addressed.

At the heart of understanding these complex systems, lies the science of modeling, as noted by \citet{Rauzy:MBSE:2019}. As these systems grow more intricate, refining the corresponding models and techniques becomes critical to maintaining an appropriate level of abstraction. This is particularly evident in Probabilistic Risk Analysis (PRA), an important field within reliability engineering and risk sciences. PRA models integrate various probabilistic models, including event trees, Bayesian networks, Markov chains, Stochastic Petri Nets, and Reliability Block Diagrams, along with Event Sequence Diagrams and their extensions. These models encapsulate system-theoretic knowledge, encompassing aspects such as system composition and behavior, as well as risk-related elements like critical failure modes and probabilities of component failures, all underpinned by underlying graph models.

In this context, achieving the right level of abstraction in modeling complex systems is essential. However, as systems and models increase in complexity the effort required to construct and solve them also increases. While there are numerous works focus on effective model generation \citet{Jimenez-Roa:Auto_FT_Generation:2023, mandelli:auto_event_fault_trees:2023}, this paper focuses on the application of recent advances in Artificial Intelligence (AI). Particularly on the synergy between Graph Neural Networks (GNNs) and Reinforcement Learning (RL) techniques, to address the challenges of solving models of increasingly complex systems.

The contribution of this paper is to present a novel conceptual framework that unites traditional PRA with modern Machine Learning (ML) approaches, enhancing our capability to solve complex models, and thus increasing our understanding of complex systems. The structure of the paper is designed to guide the reader through this journey, beginning with a brief presentation of the foundational aspects of FTs, RL, and GNNs in Section \ref{sec: Basis}, followed by the introduction of our proposed concept in Section \ref{sec: Concept}.

This paper is a working document and a preprint version. As such, it represents an evolving exploration of the concept, which we anticipate will develop further as we continue our research.


\section{Basis}\label{sec: Basis}

This section presents a brief presentation of the foundational elements of our concept.


\subsection{Fault Trees}
Fault Trees (FTs) are one of the most commonly used PRA modeling paradigms. FTs enable the identification of how different system faults (basic events) are logically connected and interact to cause a specific system failure (top event). They employ logical gates such as AND, OR, K-out-of-N, etc., to model these interactions. For example, a system might be deemed to fail if both primary and spare components fail simultaneously. The Fault Tree Analysis (FTA) is traditionally focused on the identification of so-called Cut Sets (CS), unique combinations of basic events leading to the top event, sometimes there is an emphasis on Minimal Cut Sets (MCS), the CSs with the minimum number of basic events that lead to a top event. The identification of CS is integral to understanding the likely failure pathways within a system. FTs also facilitate quantitative analysis, providing metrics such as the probability of system failure, the mean number of failures for discrete-time, the mean downtime (unavailability), Mean Time To Failure (MTTF), as well as Mean Time To Repair (MTTR), and Mean Time Between Failures (MTBF) for continuous-time Standard Fault Trees (SFT) that can model failure rates \citep{Ruijters:FTA:2015}. These metrics can be computed either directly using top-down or bottom-up methods or using underlying Binary Decision Diagrams (BDDs) models \citep{rauzy1993new}. BDD-based methods are used in many cases since they allow the computation of the system unreliability and the probabilities of the CS simultaneously. While SFTs are simple and commonly used, they cannot model dynamic behavior and account for uncertainties \citep{Ruijters:FTA:2015}. As such, several extensions have evolved such as Dynamic Fault Trees (DFTs) \cite{DFT}, Fault Trees with Dependent Events (FTDE) \cite{FTDE}, Repairable Fault Trees (RFT) \cite{RFT}, State-Event Fault Trees (SEFT) \cite{SEFT}, Component Fault Trees (CFT) \cite{CFT}, and Boolean logic-Driven Markov Processes (BDMP) \cite{BDMP}. They now form a broad class of dynamic risk assessment methods. 

DFTs extend intuitive, logical failure representation of static fault trees with additional dynamic gates such as PAND, FDEP, and SPARE, allowing for the modeling of complex behaviors, interactions, and failure sequences of system components. For example, a SPARE gate activates spare units when primary units fail, continuing until no more spares are available. Basic events in a DFT can be dormant, active, or failed, with the failure probability of dormant components reduced by the dormancy factor. Quantitative analysis of DFTs typically involves applying standard methods to their operational semantics in terms of Continuous-Time Markov Chains (CTMCs), Interactive and Input-Output Markov Chains, or dynamic Bayesian networks (BNs). 


\subsection{Reinforcement Learning}\label{sec: RL}
RL operates on a paradigm where agents continuously learn and refine their decision-making abilities through interactions with an environment. Diverging from traditional ML methods, RL is distinguished by its reliance on reward signals for directing the learning process, in the absence of pre-labeled data. A critical aspect of RL is the equilibrium between exploration, which involves experimenting with novel actions, and exploitation, which focuses on using established action-outcome relationships to make decisions. As depicted in Figure \ref{Fig: RL-Loop}, the RL training loop is an iterative cycle. In this cycle, an agent executes an action, the environment then updates the state and provides a reward, and the agent uses this feedback to inform its future actions. The essence of RL lies in the agent's continuous refinement of its policy, a process driven by the dual forces of exploration and exploitation: seeking new, potentially superior strategies while utilizing known tactics to accumulate maximum rewards.

RL represents a paradigm where agents iteratively learn optimal decision-making through interaction with an environment. Distinct from other ML methodologies, RL is uniquely characterized by the use of reward signals to guide learning, the absence of labeled data, and the pivotal need to balance exploration - trying out new actions, with exploitation - leveraging known action-consequence relationships. Figure \ref{Fig: RL-Loop} illustrates the general training loop in RL, highlighting the iterative process where an agent prescribes an action, the environment responds with a new state and reward, and this feedback informs the agent's subsequent actions. The RL method manifests as a cyclic process wherein the agent refines its policy over successive interactions with the environment. This refinement occurs through a balance of exploration—seeking new strategies to uncover optimal actions—and exploitation—leveraging known strategies to maximize cumulative rewards.

\begin{Figure}
    \centering
    \includegraphics[width=\linewidth]{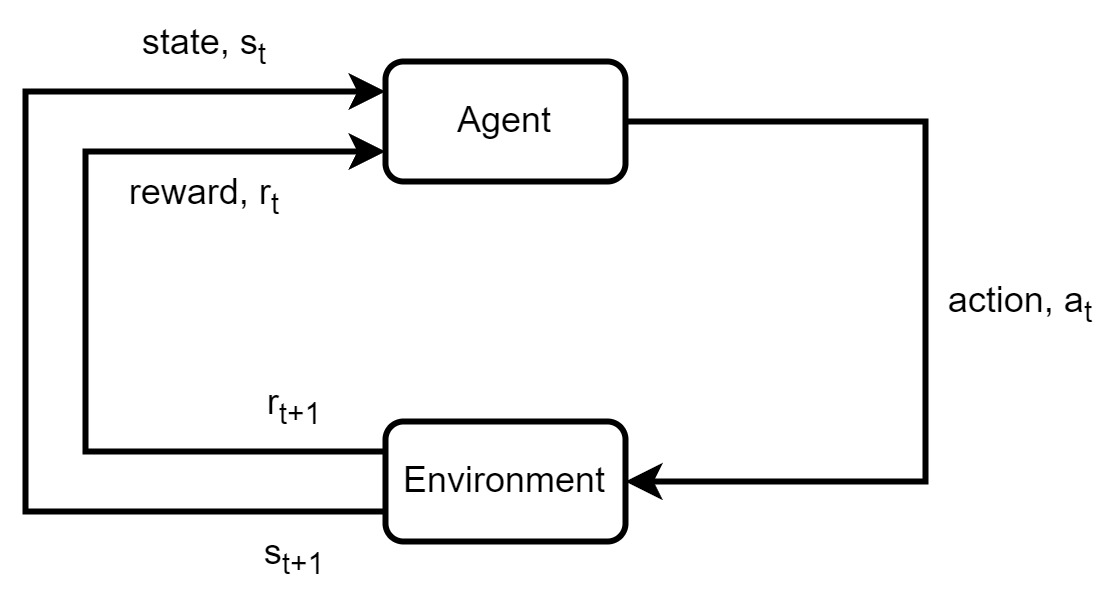}
    \captionof{figure}{The general RL training loop}
    \label{Fig: RL-Loop}
\end{Figure}


Some key concepts in RL include:

\textbf{Agent}: An algorithmic entity equipped with decision-making capabilities, tasked with navigating and interacting with a given environment.

\textbf{Environment}: A dynamic system characterized by states, actions, and rewards, encapsulating the contextual landscape in which the agent operates.

\textbf{State (s)}: State is the representation of the features defining the current situation within the environment. 

\textbf{Observation (o)}:
Sometimes it is more realistic to train a policy with incomplete information, therefore observations are sometimes used to differentiate between the state of the environment and the observations available to the agent. i.e. observations are a subset of state.

\textbf{Action (a)}: The strategic decision prescribed by the agent in response to its perceived state, influencing subsequent interactions.

\textbf{Reward (r)}: A feedback signal from the environment, serving as an evaluative measure of the immediate consequence resulting from the agent's prescribed action.

\textbf{Policy ($\pi, \mu$)}: The mapping or strategy employed by the agent to determine its actions based on observed states, can be noted as $\pi$ or $\mu$ for stochastic or deterministic policies respectively. 


\subsubsection{Proximal Policy Optimization}\label{PPO}
Proximal Policy Optimization (PPO) proposed by \citet{Schulman:PPO:2017} is a state-of-the-art reinforcement learning algorithm designed for training agents to make sequential decisions in complex environments. PPO belongs to the family of policy optimization algorithms and is specifically formulated to address stability and sample efficiency concerns. One notable feature of PPO is its emphasis on maintaining a conservative policy update strategy, preventing large policy changes during training to ensure smoother convergence by clipping the objective function. By constraining the policy changes, PPO mitigates the risk of divergent learning and enhances stability, making it well-suited for a variety of applications. Its effectiveness in balancing the trade-off between exploration and exploitation has positioned PPO as a popular choice in RL.

\subsection{Graph Neural Networks}\label{sec: GNN}
Graph Neural Networks (GNNs) are a class of neural network architectures specifically tailored for processing and analyzing graph-structured data first proposed by \citet{Scarselli:GNN:2008}. GNNs offer a powerful tool for capturing intricate relationships and dependencies within complex systems. Unlike traditional neural networks that pre-process data into vector spaces, GNNs operate on irregular and interconnected data representations. This makes them particularly suitable for applications where data can be represented in the form of graphs such as system theory and risk models.

Graphs are typically expressed as an ordered pair of sets, $G = (V, E)$ where $V$ denotes the set of vertices (often called nodes), and $E$ denotes the set of edges (often called connections). An alternative expression of graphs is as an adjacency matrix, $A_{|V| \times |V|}$, where each element, $a_{mn} \in \{0, 1, 2\}$. Here 0, represents no edge-, and 1 represents one edge between the vertices $m$ and $n$. It follows that 2 represents a self-loop.

GNN tasks are divided into three categories, these are vertex-level, edge-level, and graph-level tasks \citep{Zhou:GNN:Survey:2020, Wu:GNN:Survey:2021, Waikhom:GNN:Survey:2023}. Here vertex-level tasks include vertex classification, - regression, and - clustering. 
Classification tasks are designed to categorize vertices within a graph, effectively assigning them to specific classes based on their characteristics. On the other hand, vertex regression tasks focus on predicting the features of vertices. Meanwhile, clustering involves grouping similar vertices, creating cohesive segments within the graph. For edges, methods include edge classification and link prediction. Methods that classify edges and predict the existence of edges between vertices respectively. Classification and regression can also be done on the graph level \citep{Zhou:GNN:Survey:2020}.


\section{Concept}\label{sec: Concept}

\subsection{Motivation}\label{sec: Motivation}

The increasing complexity of modern systems, marked by their intricate interdependencies and dynamic behaviors, underscores the urgent need for advanced methodologies in PRA and the greater field of reliability- and safety engineering. While traditional approaches have demonstrated robustness, they often struggle to fully encompass and analyze the multi-layered and ever-changing nature of these systems. The challenge is further amplified by the sheer scale of these systems, often represented by a multitude of diverse models, making it difficult to gain a holistic understanding. This evolving landscape calls for innovative solutions that can integrate and make sense of the expansive and dynamic scope of system models, ensuring a comprehensive view of system reliability, safety, and risk.

\subsection{General Concept}\label{sec: General Concept}

The general concept is to develop models capable of solving specific metrics or characteristics of FTs, yet be able to generalize such that it can solve for new FTs not previously seen in training. This can then be extended in complexity or to other methods underpinned by graph models. As shown in Figure \ref{Fig: FT} and \ref{Fig: Graph}, FTs can be represented as acyclic-directed graphs where information about intermediate events, gates, probabilities, etc., can be encoded as features of the vertices and edges.       

\begin{Figure}
    \centering
    \includegraphics[width=\linewidth]{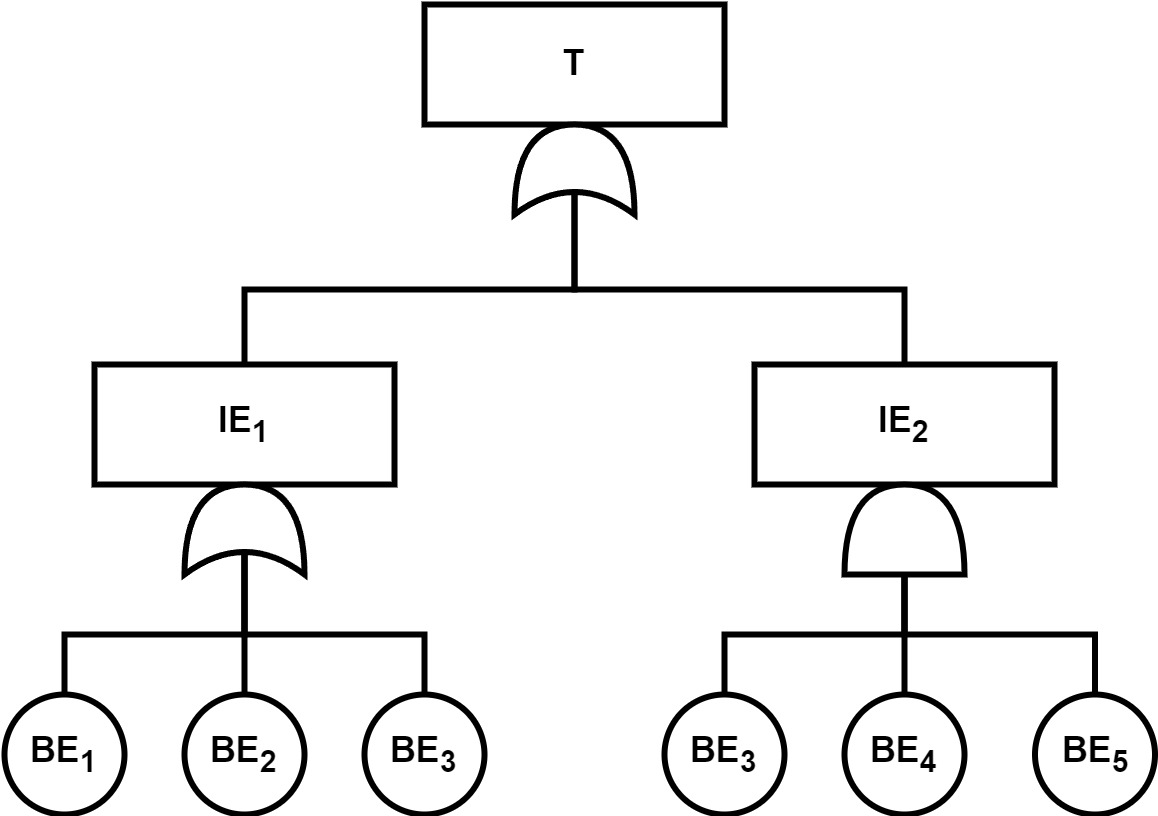}
    \captionof{figure}{Example fault tree}
    \label{Fig: FT}
\end{Figure}

\begin{Figure}
    \centering
    \includegraphics[width=\linewidth]{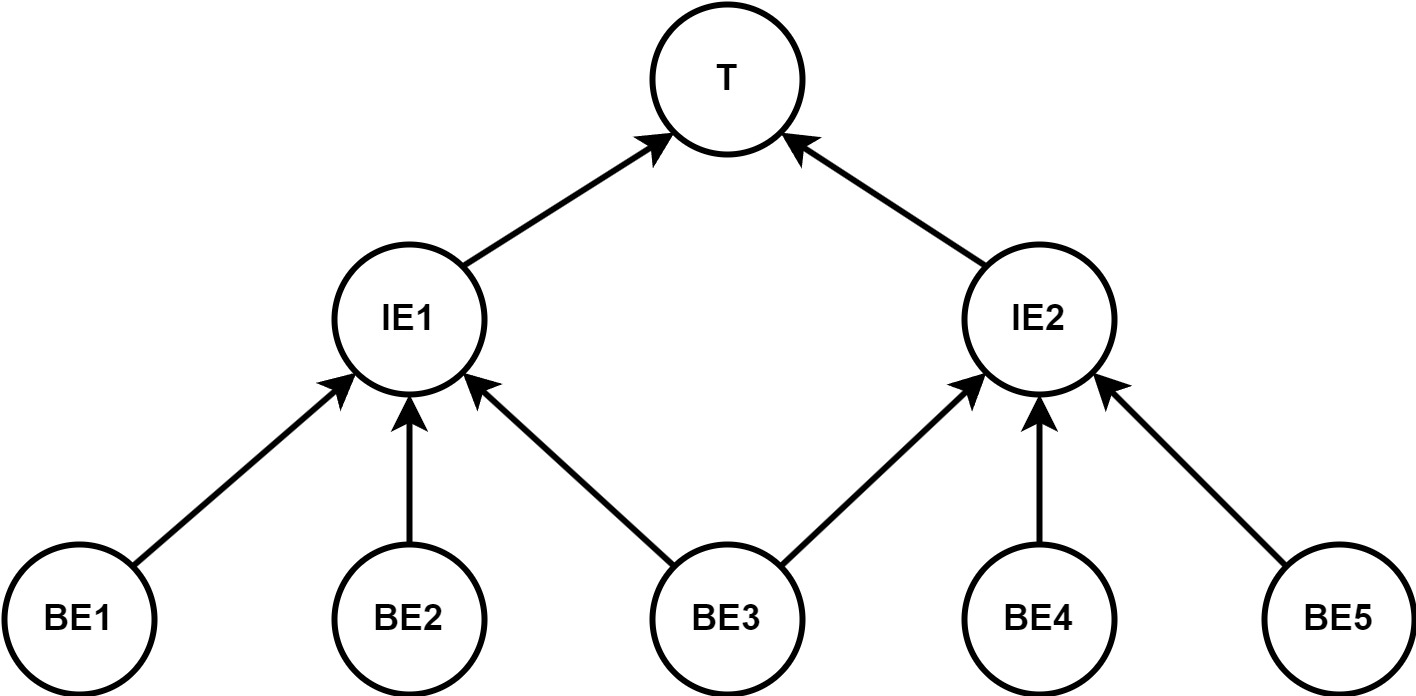}
    \captionof{figure}{Example acyclic-directed graph}
    \label{Fig: Graph}
\end{Figure}

\subsection{Vertex Level}\label{sec: Concept}

\subsubsection{Quantitative}

Quantitative analysis in FTA plays a crucial role in evaluating the reliability and safety of complex systems. By focusing on node-level features within FTs, it becomes possible to gain a nuanced understanding of how individual components influence the overall system. This detailed approach is not only informed by historical data but can also be enhanced through contemporary machine-learning techniques.

In instances where extensive data is available, robust training datasets can be developed, allowing for the application of data-driven models to prescribe node-level features accurately. These models can analyze large volumes of reliability and failure data, providing precise quantitative measures for each node in the FT.

Moreover, in situations where data might be scarce or insufficient for traditional methods, RL offers a dynamic alternative. In this approach, agents are trained to estimate the quantitative features of each node, with their performance being evaluated against the ground truth. The agents receive rewards based on the accuracy of their feature prescription, iteratively improving their capabilities through RL. This method allows for the continual refinement of quantitative assessment, adapting to new data and evolving system behaviors, thereby enhancing the overall reliability assessment process in FTA.

A simple example of how RL might be trained to prescribe the failure probabilities of a simple SFT. consider Figure \ref{Fig: RL-Loop}, the agent observes the state of the environment, including failure probabilities of basic events, and the underlying structure of the FT. Then the agent prescribes failure probabilities for the desired vertices, this is then compared to ground truth, and a reward is calculated based on the relative difference. The episode could then continue for a certain number of iterations and the agent would try to accumulate the maximum reward possible reward over an episode. Each episode would have a different FT allowing the model to generalise and learn how failure probabilities aggregate through a model with a different structure. How the rewards are designed also influences the performance and applicability of the model; One example is to reward the agent relative to the difference in the prescribed attribute $\leq$ ground truth. Conversely, punish the agent relative to the difference if prescribed attribute $>$ ground truth, thus training a more pessimistic model. 

\subsection{Edge Level}\label{sec: Edge}

\subsubsection{Dependency}\label{sec: Dependancy}

Understanding dependencies between various failure modes is a critical aspect of FTA. GNNs, with their edge/link prediction capabilities \citep{Zhang:Link_prediction:2018}, could offer a sophisticated approach to uncovering potential hidden dependencies, as well as the correlation/strength, and uncertainty of these hidden dependencies. Recent works have also focused on enhancing the scalability of link prediction in GNNs, making their training more effective and efficient, especially in large-scale systems \citep{Karunarathna:Link_scale:2020, Senevirathne:Link_prediction_training:2020}. Additionally, advancements in link prediction for evolving graphs have been explored \citep{Baek:Link_prediction_extrapolate:2020}, addressing the dynamic nature of fault trees where system configurations and failure modes may change over time.

\subsection{Graph Level}\label{sec: Tree}

One intriguing aspect is the GNN's ability to alter the input graph. As GNNs process the fault tree graph, they can be trained to not only analyze but also dynamically modify the graph structure based on learned patterns and interactions. This alteration might involve the addition or removal of vertices and edges, or the updating of vertex features, to better reflect the underlying system dynamics and interdependencies. For instance, a GNN may identify and introduce new vertices representing latent failure modes or adjust the connections between existing vertices to represent discovered dependencies more accurately. These modifications, derived from the GNN's deep learning capabilities, could potentially offer a more accurate and comprehensive representation of the system's reliability, going beyond the limitations of the original model. This could not only be used as a tool to maintain models during the life cycle of complex systems based on new information or modifications but could also be used to isolate and enhance our understanding of aspects of the graph, underlying models, and thus the complex system. One common such instance is to identify CSs, minimal, most probable, or otherwise. One simple example would be to identify MCSs. Consider Figre \ref{Fig: RL-Loop}, the agent can remove edges and nodes as actions, and the graph is then compared to the ground truth. If the new graph is an invalid graph or CS, a negative reward signal is sent, however, if the new graph is a valid CS the agent is rewarded based on the number of vertices in the CS, $(|v| - |v_{cs}|)$.

\begin{Figure}
    \centering
    \includegraphics[width=\linewidth]{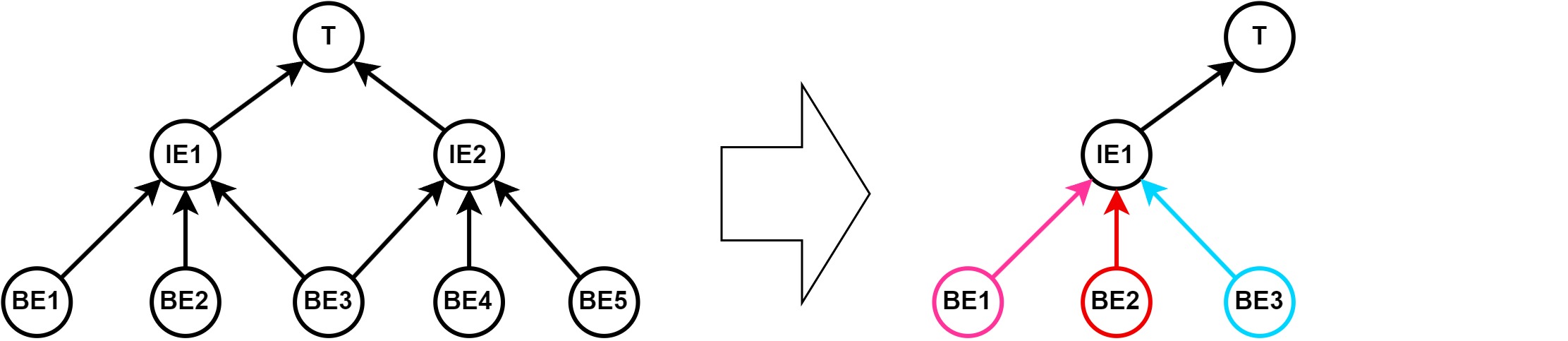}
    \captionof{figure}{example of the MCS $= (\{BE1\}, \{BE2\}, \{BE3\})$ }
    \label{Fig: mcs}
\end{Figure}

\section{Conclusion}\label{sec: Conclusion}

In conclusion, this paper has explored the application of GNNs in enhancing FTA. A concept for how GNNs can effectively analyze and modify fault trees at both node and tree levels, providing deeper insights into system failures and dependencies has been proposed. While challenges such as, model complexity, transparency, and explainability must be addressed, this approach could represent a significant step forward in reliability engineering, offering a data-driven method to improve the accuracy and efficiency of PRA. Future work should focus on refining these methods, ensuring they are practicable for use in complex system analysis.





\bibliographystyle{unsrtnat}
\bibliography{references}  

\end{multicols}
\end{document}